\documentclass[10pt,letterpaper]{article}

\usepackage[top=0.85in,left=1.75in,footskip=0.75in]{geometry}
\usepackage{changepage}
\usepackage[utf8]{inputenc}
\usepackage{textcomp,marvosym}
\usepackage{fixltx2e}
\usepackage{amsmath,amssymb}
\usepackage{cite}
\usepackage{nameref,hyperref}
\usepackage{microtype}
\DisableLigatures[f]{encoding = *, family = * }
\usepackage{rotating}

\raggedright
\usepackage[aboveskip=1pt,labelfont=bf,labelsep=period,justification=raggedright,singlelinecheck=off]{caption}

\makeatletter
\renewcommand{\@biblabel}[1]{\quad#1.}
\makeatother

\date{}

\usepackage{lastpage,fancyhdr,graphicx}
\usepackage{epstopdf}
\fancyheadoffset[L]{2.25in}
\fancyfootoffset[L]{2.25in}

\usepackage[nolist]{acronym}
\usepackage{xspace}
\usepackage[babel]{csquotes}
\usepackage[table]{xcolor}
\usepackage{subfigure}
\usepackage{snapshot}

\begin{acronym}[Bash]
 \acro{HMM}{Hidden Markov Model}
 \acro{CRF}{Conditional Random Field}
 \acro{POS}{Part-Of-Speech}
 \acro{MEMM}{Maximum Entropy Markov Model}
 \acro{LDCRF}{Latent-Dynamic Conditional Random Field}
 \acro{HCI}{Human-Computer Interaction}
 \acro{HRI}{Human-Robot Interaction}
 \acro{ROPAMS}{Remote Object Position Attitude Measurement System}
 \acro{HCRF}{Hidden Conditional Random Field}
 \acro{SVM}{Support Vector Machine}
\end{acronym}

\makeatletter
\DeclareRobustCommand\onedot{\futurelet\@let@token\@onedot}
\def\@onedot{\ifx\@let@token.\else.\null\fi\xspace}

\def\eg{\emph{e.g}\onedot} 
\def\ie{\emph{i.e}\onedot}

\def\etal{\emph{et al}\onedot}

\newcommand{\scaletables}[1]{\scalebox{0.9}{#1}}
\newcommand{\mcrot}[4]{\multicolumn{#1}{#2}{\rlap{\rotatebox{#3}{#4}~}}}

\newcommand{\disablegraphics}[1]{#1}

\begin{document}
\vspace*{0.35in}

\begin{flushleft}
{\Large
\textbf\newline{What's the point? Frame-wise Pointing Gesture Recognition with Latent-Dynamic Conditional Random Fields}
}
\newline
\\
Christian Wittner\textsuperscript{1},
Boris Schauerte\textsuperscript{1},
Rainer Stiefelhagen\textsuperscript{1}
\\
\bigskip
\bf{1} Institute for Anthropomatics and Robotics, Karlsruhe Institute of Technology, Karlsruhe, Germany\\
\bigskip

%
%





* boris.schauerte@kit.edu

\end{flushleft}
\section*{Abstract}
We use Latent-Dynamic Conditional Random Fields to perform skeleton-based pointing gesture classification at each time instance of a video sequence, where we achieve a frame-wise pointing accuracy of roughly $83\%$.
Subsequently, we determine continuous time sequences of arbitrary length that form individual pointing gestures and this way reliably detect pointing gestures at a false positive detection rate of $0.63\%$.

\section{Introduction}\label{sec:introduction}

Pointing gestures are a fundamental aspect of non-verbal human-human interaction, where they are often used to direct the conversation partner's attention towards objects and regions of interest -- an essential mean to achieve a joint focus of attention.
As a consequence, reliable detection and interpretation of pointing gestures is an important aspect of natural, intuitive \acf{HCI} and \acf{HRI}.

In this paper, we use \acfp{LDCRF} to perform pointing gesture detection and frame-wise segmentation based on skeleton data such as, \eg, the joint data provided by a Kinect.
Therefore, we use the \ac{LDCRF} to label each frame of a video sequence and subsequently determine continuous time sequences of arbitrary length that form individual pointing gestures.
An important advantage of this approach is that we can detect pointing gestures while they are being performed and do not have to wait for a person to perform and complete the whole pointing action.
This enables us to react to a pointing gesture as it is performed -- an important aspect considering that natural \ac{HRI} is our target application (see, \eg, \cite{schauerte2014look,schauerte2010focusing}).
For example, this way our robot is able to direct its head toward the coarse target area, thus providing a visual feedback for the pointing person, while the pointing person is still adjusting and/or fine-tuning the pointing direction.
We evaluate the performance of our pointing gesture detection method based on a novel dataset. 
We used a Microsoft Kinect to record a diverse set of pointing gestures performed by 18 subjects, which enables us to evaluate person-independent pointing gesture detection.

\newlength{\exampleheightobama}
\setlength{\exampleheightobama}{2.125cm}
\begin{figure}[tb]
  \disablegraphics{%
	  \centering
	  \includegraphics[width=.975\linewidth]{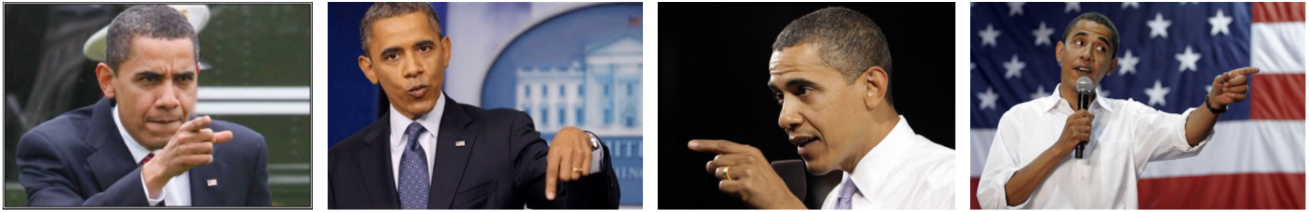}
  }
  \caption{Pointing gestures are an essential aspect of human communication that is frequently used throughout various situations such as, e.g., during speeches.}
\end{figure}


\section{Related Work}\label{sec:relwork}

%
%
%

One of the first systems to recognize pointing gestures was Bolt's ``Put that there'' system \cite{Bolt1980}, which enabled users to point at and interact with objects while giving voice commands. 
However, the system required the user to wear a Polhemus \acf{ROPAMS} device at his wrist. 
Systems that avoid specialized wearable devices nowadays mostly use the data provided by stereo or depth cameras as a basis for (pointing) gesture recognition.
Here, stereo cameras or depth cameras (\eg, Microsoft Kinect or the Mesa SwissRanger) are used to acquire depth image sequences, which allows for simpler gesture recognition in situations with, for example, occlusions and/or difficult lighting conditions.

Gesture recognition for \acl{HCI} has been an active research areas for many years. 
Accordingly, there exist several surveys that provide a detailed overview of early and recent research developments 
\cite{Suarez2012,Mitra2007,Wachs2011,Gavrila1999,Pavlovic1997,Jaimes2007}.
In the following, we focus on depth-based sequential gesture recognition approaches and, for example, do not discuss non-sequential approaches (\eg, \cite{Jojic2000,Feris2005,VanDenBergh2011,Keskin2011,Biswas2011,Ramey2011}) 

\acfp{HMM}
and stochastic grammars are widely used models for speech and gesture recognition. 
The Hidden Markov Model is a generative model that includes hidden state structure. 
Gesture recognition applications using \acp{HMM} can be found at Chen \etal \cite{chen2003hand}, Yang \etal \cite{yang2012gesture}, Zafrulla \etal \cite{zafrulla2011american} and Vogler \etal \cite{vogler1998asl}.
The popular approach by Nickel and Stiefelhagen \cite{Nickel2003} uses a stereo camera to detect pointing gestures. 
A color-based detection of hands and face is combined with a clustering of the found regions based on depth information enabling the tracking of 3D skin color clusters. 
Hidden Markov Models are trained on different states (begin, hold, end) of sample pointing gestures and used to detect the occurrence of pointing gestures. 
The additional head orientation feature is used to improve recall and precision of pointing gestures. 
The authors claim to achieve 90\% accuracy identifying targets using the head-hand line, but no accuracy for the pointing gesture detection has been reported.
In Park \etal \cite{park2008real}, face and hand tracking is performed using 3D particle filters after initially having detected those body parts based on skin color. 
To account for small pointing gestures (where the pointing arm is not fully extended), the hand positions are mapped onto an imaginary hemisphere centered around the shoulder of the pointing arm before estimating the pointing direction. 
This is done by a first stage \ac{HMM} that is used to retrieve more accurate hand positions. 
In a second stage these hand positions are fed into three \acp{HMM} (for three states Non-Gesture, Move-To and Point-To) in order to detect a pointing gesture. 
In case of a pointing gesture, the pointing direction is estimated.

Droeschel \etal \cite{droeschel2011learning} focus on pointing gestures where the person does not look into the target direction. 
Body features are extracted from the depth and amplitude images delivered by a Time-of-Flight camera. 
\acp{HMM} are used to detect a pointing gesture which is segmented into the three phases \enquote{preparation}, \enquote{hold} and \enquote{retraction}. 
The \acp{HMM} are trained with features such as the distance from the head to the hand, the angle between the arm and the vertical body and the velocity of the hand. 
To estimate pointing directions, a pointing direction model is trained using Gaussian Process Regression, 
which leads to a better accuracy than simple criteria like head-hand, shoulder-hand or elbow-hand lines (see \cite{Nickel2003}).

Despite their popularity, \acp{HMM} have some considerable limitations. 
An \ac{HMM} is a generative model that assumes joint probability over observation and label sequences. 
The model needs to enumerate all possible observation sequences in which each observation is required to be an atomic entity. 
Thus, the representation of long-range dependencies between observations or interacting features is computationally not tractable.

The need for a richer representation of observations (\eg, with overlapping features) has led to the development of \acfp{MEMM} \cite{mccallum2000maximum}.
A \ac{MEMM} is a model for sequence labeling that combines features of \acp{HMM} and Maximum Entropy models. 
It is a discriminative model that extends a standard maximum entropy classifier by assuming that the values to be learned are connected in a Markov chain rather than being conditionally independent of each other. 
It represents the probability of reaching a state given an observation and the previous state.
Sung \etal \cite{sung2012unstructured} implemented detection and recognition of unstructured human activity in unstructured environments using hierarchical \acp{MEMM}. 
Sung \etal use different features describing body pose, hand position and motion extracted from a skeletal representation of the observed person. 

A major shortcoming of \acp{MEMM} and other discriminative Markov models based on directed graphical models is the \enquote{label bias problem} which can lead to poorer performances compared to \acp{HMM}. 
This problem has been addressed by Lafferty \etal's \acfp{CRF} \cite{lafferty2001conditional}.
A conditional model like \acp{CRF} specifies the probabilities of label sequences for a given observation sequence. 
Features of the observation sequence do not need to be independent and they may represent attributes at different levels of granularity of the same observations. 
They could combine several properties of the same observation. 
Past and future observations may be considered to determine the probability of transitions between labels.
Lafferty \etal \cite{lafferty2001conditional} and Sminchisescu \etal \cite{sminchisescu2006conditional} demonstrate how \acp{CRF} outperform both \acp{HMM} and \acp{MEMM}. 
While Lafferty \etal \cite{lafferty2001conditional} use synthetic data and real \acf{POS} tagging data, Sminchisescu \etal \cite{sminchisescu2006conditional} apply \acp{CRF} for the recognition of human motions.
For this purpose,
Sminchisescu \etal use a combination of 2D features from image silhouettes and 3D features. 
They show that the \ac{CRF}'s performance based on 3D joint angle features is more accurate with long range dependencies and that \acp{CRF} improve the recognition performance over \acp{MEMM} and \acp{HMM}.

\acp{CRF} can model the transitions between gestures (extrinsic dynamics) but are not able to represent internal sub-structure. 
This led to the development of Hidden Conditional Random Fields that incorporate hidden states variables that model the sub-structure of a gesture sequence.
Wang \etal \cite{Wang2006} combine the two main advantages of current approaches to gesture recognition: The ability of \acp{CRF} to use long range dependencies and the ability of \acp{HMM} to model latent structure. 
One single joint model is trained for all gestures to be classified and hidden states are shared between those gestures. 
According to Wang \etal \cite{Wang2006} the \acfp{HCRF} model outperforms \acp{HMM} and \acp{CRF} in the classification of arm gestures.
But since \acp{HCRF} are trained on pre-segmented sequences of data, they only capture the internal structure but not the dynamics between gesture labels. 
To overcome this limitation, Morency \etal \cite{morency2007latent} introduced \acfp{LDCRF}, 
that combine the strengths of \acp{CRF} and \acp{HCRF}. 
\acp{LDCRF} are able to capture extrinsic dynamics as well as intrinsic substructure and can operate directly on unsegmented data. 
The performance of \acp{LDCRF} was tested on head and eye gesture data of three different datasets and the results were compared to state-of-the-art generative and discriminative modeling techniques like \ac{SVM}, \acp{HMM}, \acp{CRF} and \acp{HCRF}. 
The results show that \acp{LDCRF} perform better than all other methods.


\section{Method}\label{sec:method}

In the following, we use the Microsoft Kinect and OpenNI's NITE \cite{shotton2013real} module to obtain a skeleton of the person in front of the Kinect.
We then use the joint data over time to detect the occurrence of pointing gestures.
Since the Kinect provides relatively stable joint tracks (\eg, it is quite robust against illumination changes), we focus on pointing gesture detection and do not address, \eg, noisy body part detections.
However, our trained model can be applied to other sensing methods (\eg, stereo cameras), if comparable, stable joint tracks are provided.

\subsection{CRF}

\acfp{CRF} are a framework for building probabilistic models to segment and label sequence data \cite{lafferty2001conditional}. 
%
%
For this purpose,
\acp{CRF} define an undirected graph $G=(V,E)$ with the random variable $X$ representing the input variable observation and $Y$  being a random variable over corresponding label sequences. 
All components $Y_v \in Y$ are from the label alphabet $\mathcal{Y}$ = \{\enquote{background}, \enquote{rise left}, \enquote{point left}, \enquote{fall left}, \enquote{rise right}, \enquote{point right}, \enquote{fall right}, \enquote{other}\}. $G$ contains a vertex $v \in V$ for each random variable representing an element $Y_v \in Y$. 

%
%

In contrast to \acp{HMM}, \acp{CRF} do not model the output label sequence $Y$ and input data sequence $X$ as a joint probability $P(X,Y)$, but the conditional probability of a label sequence $Y_j$ given a sequence of input data $X_i$ and the model parameters $\theta$ 
\begin{equation} \label{eq:crf_label_probabilities}
P(Y_j|X_i; \theta) = \frac{1}{Z(X_i, \theta)} exp \left( \sum_k \theta_k F_k(Y_j,X_i) \right),
\end{equation}
with the normalizing partition function 
\begin{equation}
Z(X_i, \theta) = \sum\limits_j exp \left(  \sum_k \theta_k F_k(Y_j,X_i) \right)
\end{equation}
summing for each $X_i$ over all $Y_j$ corresponding with $X_i$.

The feature functions $F_k (Y_j , X_i )$ can be broken down into state functions $s_l(y_f,X_i,f)$ and transition functions $t_m(y_f,y_{f-1},X_i,f)$
\begin{equation} \label{eq:crf_feature_functions}
F_k(Y_j,X_i) = \sum_f \left( \sum\limits_{l} \lambda_l s_l(y_f,X_i,f) + \sum_{m}  \mu_m t_m(y_f,y_{f-1},X_i,f) \right).
\end{equation}

State functions $s_l$ model the conditioning of the labels $Y$ on the input features $X$ while the transition functions $t_m$  model the relations between the labels $Y$ with respect to input features $X$. 
Due to computational constrains a transition feature functions can only have two label node values as input parameters.

\begin{figure}[tb]
        \centering
        \disablegraphics{%
                \includegraphics[width=.5\linewidth]{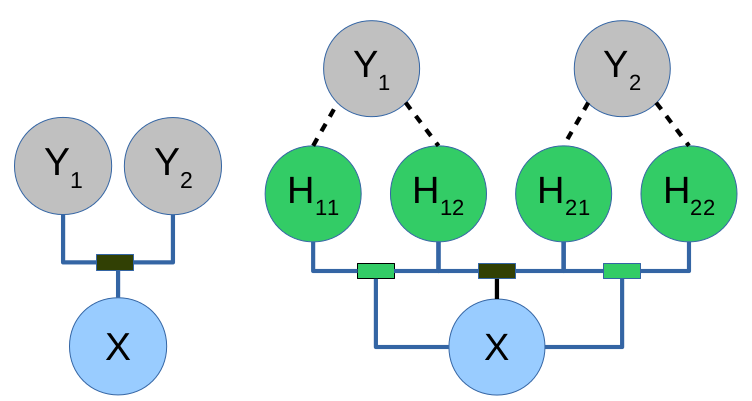}
                
        }
        \caption{\ac{CRF} (left) and \ac{LDCRF} (right) structure illustration with labels $Y_v$, features $X$, and latent variables $H_{ij}$. The boxes represent transition functions (green: intrinsic; black: extrinsic).}\label{fig:random_fields}\label{fig:crf}\label{fig:ldcrf}
\end{figure}

For our approach, the unconstrained access to any input feature $x_f \in X_i$, before or after a given frame, in the sequence $X_i$, is one of the major advantages that \acp{CRF} provide in contrast to \acp{HMM}
(see Sec. \ref{sec:feature_history}). 
 
\subsection{LDCRF}

\acfp{LDCRF} as introduced by Morency \etal \cite{morency2007latent} (please note that \acp{LDCRF} were first introduced as Frame-based Hidden-state Conditional Field in Morency's PhD-thesis \cite{morency2006context}), extend the structure of \acp{CRF} by hidden states to model intrinsic structure.
Therefore, a set of latent variables $\mathcal{H}$ is introduced. 
Here, three hidden states per label $\mathcal{Y}$.
%
The probability of each label in the graph is substituted by the chain of probabilities of its hidden states
\begin{equation} \label{eq:ldcrf_model_probability_long}
P(Y|X;\theta) = \sum_{H \in \mathcal{H}} P(Y|H,X;\theta) P(H|X;\theta).
\end{equation}

The random field (see Eq. \ref{eq:crf_label_probabilities}) is now build upon the hidden states $\mathcal{H}$
\begin{equation} 
P(H_n|X_i; \theta) = \frac{1}{Z(X_i, \theta)} exp \left( \sum_k \theta_k F_k(H_n,X_i) \right).
\end{equation}

Similar to \acp{CRF}, transition functions model relations between two hidden states. 
As shown in Fig.~\ref{fig:random_fields} these functions can in addition to the extrinsic structure (black boxes), relating observable class nodes, now also model intrinsic structure of a class (green boxes). 

Interestingly, \acp{HMM} model intrinsic structure with hidden states as well, but they need a model for each label class $\mathcal{Y}$.
Thus, they calculate a probability for each sequence from each trained \ac{HMM} model, but those probabilities are unrelated. 
In contrast, \acp{LDCRF} seemingly combine those and output a meaningful probability for each $Y$ (\ie, $\sum_i P(Y_i) = 1$).

\subsection{Features}\label{sec:features}

Many features have been proposed and evaluated for gesture recognition (see, \eg, \cite{richarz2010feature}). 
In preliminary experiments, we evaluated several features and feature combinations (\eg, the features that Nickel and Stiefelhagen proposed \cite{Nickel2003})
We achieved the best results for pointing gesture recognition based on \acp{LDCRF} and NITE's \cite{shotton2013real} head, shoulder, elbow, and hand data with the following feature combination:
%
The
torso relative position of shoulder, elbow and hand, 
the angle between the vertical y-axis and the shoulder-hand line, 
the height difference of the two hands, 
and 
the distance of the hand to the corresponding shoulder representing the arm extension. 
To abstract different body dimensions, we normalize the distance of each distance-based feature by the shoulder width.
Furthermore, to have a completely angular representation independent of body dimensions, we add each hand's polar coordinates 
as feature.
The position of each hand in polar coordinates is represented by the azimuth angle $\alpha$ (in floor plane) and the elevation angle $\beta$ with respect to the corresponding shoulder.
Here, we only use the angles and omit the radius to be body size independent.

\subsection{History} \label{sec:feature_history}

We use the CRF's ability to establish long range relations between input frames, which is the result of each feature function's ability to access all $x_f \in X_i$ of the input sequence $X_i$ (see Eq.\,\eqref{eq:crf_feature_functions}). 
We choose to establish state functions $s_l(y_f,X_i,f)$ that use the input features $x \in \mathrm{H_{istory}} = \{x_{f-1}, x_{f-3}, x_{f-5}, x_{f-7}, x_{f-13} \}$, where all of our features described in Sec.~\ref{sec:features} are used as state functions $s_l(y_f,X_i,f)$.
This way we can represent dynamics in feature changes over an extended period.
This is useful, because, for example, at the beginning of a pointing sequence the arm rises faster while it slows down shortly before arriving at the pointing posture. 
Taking into account deltas over shorter as well as larger time intervals captures this dynamic.
We use a non-equidistant history to maintain the same level of computation complexity while covering a larger dynamic range 
compared to an equidistant history 
(\eg, $x \in \mathrm{H_{istory}} = \{x_{f-2}, x_{f-1}, x_{f}, x_{f+1}, x_{f+2} \}$ \cite{morency2007latent}).
Furthermore, we only make use of previous features (\ie, only $f+i$ with $i < 0$) since one of our system's design goals is to keep the latency as low as possible.
Thus, in contrast to, \eg, Morency \cite{morency2007latent}, we predict the current label only based on previous observations not waiting for additional future input frames.


\section{Evaluation}\label{sec:evaluation}

\newcommand{\headRotation}{30}

\subsection{Dataset}\label{sec:evaluation:dataset}

\begin{figure}
	\disablegraphics{
		\centering
		\includegraphics[width=.75\linewidth]{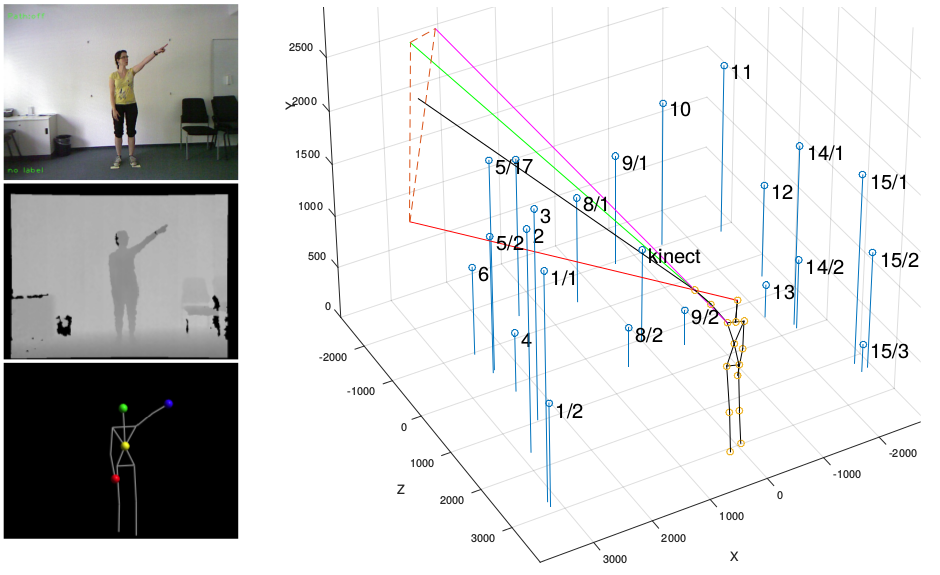}
	}
	\caption{Example of a person pointing at target 3. Left (top to bottom): The camera image, the depth image, and the skeleton. Right: The skeleton and target object locations. The different lines represent different methods to calculate the pointing gestures direction such as, \eg, the frequently used head-hand line.}\label{fig:skeleton_room}
\end{figure}

We recorded a novel evaluation dataset that consists of 990 pointing gestures: 18 persons (age range 23 to 63; six female and twelve male) performed 55 pointing gestures toward 22 targets.
The subjects were positioned approximately $4$m away from the Kinect, which is at the upper end of Kinect's depth resolution sweet spot and allows us to record a full skeleton even of the tallest subjects.
For each camera frame, we recorded the 11 bit depth image, the RGB image, and a 15 joint skeleton as is provided by the NITE framework, see Fig.~\ref{fig:skeleton_room}.
Apart from the pointing gestures, all subjects performed 10 other gestures (\eg, \enquote{hand waving}, \enquote{shoulder shrug}, or \enquote{come closer}) that we use as other/negative samples for training and testing.
Additionally, we determined each person's dominant eye.

We manually assigned every video frame in the dataset with one of 8 labels:
\enquote{background} describes idle phases between gestures (\eg, standing upright with hanging arms).
\enquote{left rise}, \enquote{left point}, and \enquote{left fall} (analogously, \enquote{right rise}, \enquote{right point}, and \enquote{right fall}) describe the typical pointing behavior of first raising the arm, the actually pointing and fine-tuning of the gesture, and finally lowering the arm again. 
\enquote{other} is used as a label for the various other gestures. 

\subsection{Results and Discussion}\label{sec:evaluation:results}

\subsubsection{Frame-wise classification}\label{sec:evaluation:results:frame}

\begin{table}[tb]
  \caption{Leave-one-subject-out evaluation results with LDCRF and CRF.\label{tab:results:frames}}
  \centering
  \subfigure[LDCRF\label{tab:leave-out-subject-trained-with-18-subjects}]{
    \centering
      \scaletables{
      \begin{tabular}{|l| c | c | c | c | c | c | c | c | c |}
      \multicolumn{1}{l}{ } & \mcrot{1}{l}{\headRotation}{Background} & \mcrot{1}{l}{\headRotation}{Left Rise} & \mcrot{1}{l}{\headRotation}{Left Point} & \mcrot{1}{l}{\headRotation}{Left Fall} & \mcrot{1}{l}{\headRotation}{Right Rise} & \mcrot{1}{l}{\headRotation}{Right Point} & \mcrot{1}{l}{\headRotation}{Right Fall} & \mcrot{1}{l}{\headRotation}{Other}\\ 
       \hline 
      Background & \cellcolor[gray]{0.17} \color{white} 83.44&\cellcolor[gray]{0.97}2.67&\cellcolor[gray]{1.00}0.43&\cellcolor[gray]{0.97}3.29&\cellcolor[gray]{0.98}2.08&\cellcolor[gray]{1.00}0.26&\cellcolor[gray]{0.98}1.83&\cellcolor[gray]{0.94}5.99\\ 
      \hline 
      Left Rise & \cellcolor[gray]{0.94}6.34&\cellcolor[gray]{0.22} \color{white} 78.14&\cellcolor[gray]{0.90}9.60&\cellcolor[gray]{0.99}1.43&\cellcolor[gray]{1.00}0.48&\cellcolor[gray]{1.00}0.00&\cellcolor[gray]{1.00}0.03&\cellcolor[gray]{0.96}3.99\\ 
      \hline 
      Left Point & \cellcolor[gray]{1.00}0.03&\cellcolor[gray]{0.96}4.28&\cellcolor[gray]{0.13} \color{white}  \textbf{\underline{87.22}}&\cellcolor[gray]{0.93}7.09&\cellcolor[gray]{1.00}0.36&\cellcolor[gray]{1.00}0.12&\cellcolor[gray]{1.00}0.15&\cellcolor[gray]{0.99}0.76\\ 
      \hline 
      Left Fall & \cellcolor[gray]{0.94}5.66&\cellcolor[gray]{0.99}0.52&\cellcolor[gray]{0.95}5.34&\cellcolor[gray]{0.15} \color{white} 85.15&\cellcolor[gray]{1.00}0.07&\cellcolor[gray]{1.00}0.11&\cellcolor[gray]{1.00}0.05&\cellcolor[gray]{0.97}3.11\\ 
      \hline 
      Right Rise & \cellcolor[gray]{0.97}2.98&\cellcolor[gray]{0.98}1.83&\cellcolor[gray]{1.00}0.07&\cellcolor[gray]{1.00}0.23&\cellcolor[gray]{0.21} \color{white} 79.02&\cellcolor[gray]{0.90}10.03&\cellcolor[gray]{0.99}0.60&\cellcolor[gray]{0.95}5.24\\ 
      \hline 
      Right Point & \cellcolor[gray]{1.00}0.00&\cellcolor[gray]{0.99}1.12&\cellcolor[gray]{0.99}0.74&\cellcolor[gray]{0.99}0.53&\cellcolor[gray]{0.91}8.79&\cellcolor[gray]{0.20} \color{white}  \textbf{\underline{79.55}}&\cellcolor[gray]{0.95}5.08&\cellcolor[gray]{0.96}4.18\\ 
      \hline 
      Right Fall & \cellcolor[gray]{0.94}6.36&\cellcolor[gray]{1.00}0.30&\cellcolor[gray]{1.00}0.08&\cellcolor[gray]{1.00}0.07&\cellcolor[gray]{0.99}0.94&\cellcolor[gray]{0.95}5.49&\cellcolor[gray]{0.20} \color{white} 80.20&\cellcolor[gray]{0.93}6.55\\ 
      \hline 
      Other & \cellcolor[gray]{0.95}5.22&\cellcolor[gray]{0.96}3.96&\cellcolor[gray]{0.98}1.53&\cellcolor[gray]{0.97}3.12&\cellcolor[gray]{0.98}2.27&\cellcolor[gray]{0.97}3.16&\cellcolor[gray]{0.98}2.23&\cellcolor[gray]{0.21} \color{white} 78.51\\ 
      \hline 
      \end{tabular}
      }
  }
  \subfigure[CRF\label{tab:one-hidden-state-simulating-crf}]{
    \centering
      \scaletables{
      \begin{tabular}{|l| c | c | c | c | c | c | c | c | c |}
      \multicolumn{1}{l}{ } & \mcrot{1}{l}{\headRotation}{Background} & \mcrot{1}{l}{\headRotation}{Left Rise} & \mcrot{1}{l}{\headRotation}{Left Point} & \mcrot{1}{l}{\headRotation}{Left Fall} & \mcrot{1}{l}{\headRotation}{Right Rise} & \mcrot{1}{l}{\headRotation}{Right Point} & \mcrot{1}{l}{\headRotation}{Right Fall} & \mcrot{1}{l}{\headRotation}{Other}\\ 
       \hline 
      Background & \cellcolor[gray]{0.25} \color{white} 74.92&\cellcolor[gray]{0.99}0.61&\cellcolor[gray]{0.93}6.91&\cellcolor[gray]{0.98}1.61&\cellcolor[gray]{0.99}0.86&\cellcolor[gray]{0.92}7.97&\cellcolor[gray]{0.99}1.26&\cellcolor[gray]{0.94}5.85\\ 
      \hline 
      Left Rise & \cellcolor[gray]{0.94}5.81&\cellcolor[gray]{0.18} \color{white} 82.01&\cellcolor[gray]{0.93}7.38&\cellcolor[gray]{1.00}0.33&\cellcolor[gray]{1.00}0.30&\cellcolor[gray]{0.99}0.99&\cellcolor[gray]{1.00}0.50&\cellcolor[gray]{0.97}2.69\\ 
      \hline 
      Left Point & \cellcolor[gray]{0.90}10.10&\cellcolor[gray]{0.97}3.22&\cellcolor[gray]{0.28} \color{white}  \textbf{\underline{71.85}}&\cellcolor[gray]{0.99}1.40&\cellcolor[gray]{1.00}0.15&\cellcolor[gray]{0.94}5.81&\cellcolor[gray]{1.00}0.13&\cellcolor[gray]{0.93}7.34\\ 
      \hline 
      Left Fall & \cellcolor[gray]{0.92}8.29&\cellcolor[gray]{1.00}0.00&\cellcolor[gray]{0.96}4.46&\cellcolor[gray]{0.15} \color{white} 85.23&\cellcolor[gray]{0.99}0.78&\cellcolor[gray]{0.99}0.56&\cellcolor[gray]{1.00}0.28&\cellcolor[gray]{1.00}0.40\\ 
      \hline 
      Right Rise & \cellcolor[gray]{0.97}3.06&\cellcolor[gray]{1.00}0.00&\cellcolor[gray]{1.00}0.27&\cellcolor[gray]{0.99}0.97&\cellcolor[gray]{0.11} \color{white} 89.16&\cellcolor[gray]{0.94}6.20&\cellcolor[gray]{1.00}0.00&\cellcolor[gray]{1.00}0.33\\ 
      \hline 
      Right Point & \cellcolor[gray]{0.84}15.52&\cellcolor[gray]{0.98}1.64&\cellcolor[gray]{0.95}5.19&\cellcolor[gray]{0.99}1.16&\cellcolor[gray]{0.97}2.69&\cellcolor[gray]{0.34} \color{white}  \textbf{\underline{65.70}}&\cellcolor[gray]{0.99}0.86&\cellcolor[gray]{0.93}7.23\\ 
      \hline 
      Right Fall & \cellcolor[gray]{0.92}7.89&\cellcolor[gray]{0.99}1.05&\cellcolor[gray]{1.00}0.33&\cellcolor[gray]{1.00}0.03&\cellcolor[gray]{1.00}0.08&\cellcolor[gray]{0.96}3.62&\cellcolor[gray]{0.13} \color{white} 86.91&\cellcolor[gray]{1.00}0.10\\ 
      \hline 
      Other & \cellcolor[gray]{0.82}18.24&\cellcolor[gray]{0.92}8.17&\cellcolor[gray]{0.78}22.17&\cellcolor[gray]{0.91}8.62&\cellcolor[gray]{0.95}5.23&\cellcolor[gray]{0.85}14.58&\cellcolor[gray]{0.95}4.63&\cellcolor[gray]{0.82}18.35\\ 
      \hline 
      \end{tabular}
      }
    }
\end{table}

\newcommand{\nanresult}{--}
\begin{table}[tb]
  \caption{Leave-one-subject-out evaluation results with HMM \cite{Nickel2003}. Please note that Nickel and Stiefelhagen's approach \cite{Nickel2003} is not multiclass and does not distinguish between left/right pointing. Accordingly, we do not record mistakes of, \eg, a detected \enquote{left rise} on a sequence in which the person points with his/her right arm.\label{tab:results:nickel}}
  \centering
  \scaletables{
  \begin{tabular}{|l| c | c | c | c | c | c | c | c |}
  \multicolumn{1}{l}{ } & \mcrot{1}{l}{\headRotation}{Background} & \mcrot{1}{l}{\headRotation}{Left Rise} & \mcrot{1}{l}{\headRotation}{Left Point} & \mcrot{1}{l}{\headRotation}{Left Fall} & \mcrot{1}{l}{\headRotation}{Right Rise} & \mcrot{1}{l}{\headRotation}{Right Point} & \mcrot{1}{l}{\headRotation}{Right Fall} & \mcrot{1}{l}{\headRotation}{Other}\\ 
   \hline 
  Background & \cellcolor[gray]{0.76}23.52&\cellcolor[gray]{0.84}15.65&\cellcolor[gray]{1.00}0.05&\cellcolor[gray]{0.95}4.82&\cellcolor[gray]{0.79}20.82&\cellcolor[gray]{1.00}0.00&\cellcolor[gray]{0.95}5.43&\cellcolor[gray]{0.70}29.71\\ 
  \hline 
  Left Rise & \cellcolor[gray]{0.99}0.78&\cellcolor[gray]{0.01} \color{white} 98.56&\cellcolor[gray]{0.99}0.66&\cellcolor[gray]{1.00}0.00& \nanresult & \nanresult & \nanresult &\cellcolor[gray]{1.00}0.00\\ 
  \hline 
  Left Point & \cellcolor[gray]{1.00}0.24&\cellcolor[gray]{0.91}8.86&\cellcolor[gray]{0.20} \color{white}  \textbf{\underline{80.33}}&\cellcolor[gray]{0.89}10.57& \nanresult & \nanresult & \nanresult &\cellcolor[gray]{1.00}0.00\\ 
  \hline 
  Left Fall & \cellcolor[gray]{0.97}2.61&\cellcolor[gray]{1.00}0.38&\cellcolor[gray]{1.00}0.07&\cellcolor[gray]{0.03} \color{white} 96.94& \nanresult & \nanresult & \nanresult &\cellcolor[gray]{1.00}0.00\\ 
  \hline 
  Right Rise & \cellcolor[gray]{0.97}2.61& \nanresult & \nanresult & \nanresult &\cellcolor[gray]{0.05} \color{white} 95.40&\cellcolor[gray]{0.99}0.60&\cellcolor[gray]{1.00}0.49&\cellcolor[gray]{0.99}0.89\\ 
  \hline 
  Right Point & \cellcolor[gray]{0.99}0.56& \nanresult & \nanresult & \nanresult &\cellcolor[gray]{0.91}9.08&\cellcolor[gray]{0.21} \color{white}  \textbf{\underline{78.81}}&\cellcolor[gray]{0.89}11.49&\cellcolor[gray]{1.00}0.06\\ 
  \hline 
  Right Fall & \cellcolor[gray]{0.93}7.35& \nanresult & \nanresult & \nanresult &\cellcolor[gray]{1.00}0.16&\cellcolor[gray]{1.00}0.20&\cellcolor[gray]{0.08} \color{white} 92.13&\cellcolor[gray]{1.00}0.16\\ 
  \hline 
  Other & \cellcolor[gray]{0.36} \color{white} 63.81&\cellcolor[gray]{0.92}8.11&\cellcolor[gray]{0.97}2.74&\cellcolor[gray]{0.95}5.27&\cellcolor[gray]{0.93}7.41&\cellcolor[gray]{0.97}3.08&\cellcolor[gray]{0.96}3.85&\cellcolor[gray]{0.94}5.73\\ 
  \hline 
  \end{tabular}
  }
\end{table}

The frame-wise classification results are depicted in confusion matrices in Tab.~\ref{tab:results:frames}.
On average, the LDCRF correctly classifies roughly $83\%$ of the frames labeled as \enquote{left/right point} that mark the holding phase with pointing on target of each gesture, see Tab.~\ref{tab:leave-out-subject-trained-with-18-subjects}.
It is important to note that the most common misclassifications are:
First, \enquote{rise} and \enquote{fall} are misclassified as either \enquote{background} or \enquote{point}.
Second, \enquote{point} is misclassified as either \enquote{rise} or \enquote{fall}.
In our opinion, such mistakes are not critical, because during the transition phases from one state into the other (\eg, from \enquote{rise} to \enquote{point}) there is substantial label ambiguity over several frames between these classes.
And these misclassifications almost exclusively occur during these transition phases.

Furthermore, we can see that the LDCRF provides a substantially better performance than the CRF, compare Tab.~\ref{tab:leave-out-subject-trained-with-18-subjects} and Tab.~\ref{tab:one-hidden-state-simulating-crf}.
Most importantly, we can see that the CRF often misclassifies frames from \enquote{other} gestures as being parts of pointing gestures, which in practice could lead to a significant amount of false positive pointing gesture misdetections.
This is an important aspect for our intended use in \acl{HRI}, because it is less disturbing for an interaction partner to repeat a pointing gesture than to have the robot react to a falsely detected pointing gesture.
The fact that the LDCRF rarely makes such mistakes is most likely due to its ability to learn and model the intrinsic structure of pointing gesture phases.

\subsubsection{Sequence detection and segmentation}\label{sec:evaluation:results:sequence}

\begin{table}[tb]
	\caption{Sequence detection results (in \%).\label{tab:prediction_state_classes}}
	\centering
	{\small
	\begin{tabular}{|l|l|r|r|}
	\hline
	Prediction Class & Type & LDCRF & HMM  \\
	\hline
	\hline
	Exact (End)                   & TP & 20.32 &  0.21 \\
	Detection longer              & TP & 23.89 &  0.31 \\ 
	Detection shorter             & TP & 29.26 & 67.26 \\
	No detection (control)        & TN &  9.47 & 10.53 \\
	\hline
	Phantom detection             & FP &  0.63 &  0.83 \\ 
	Overlapping detection         & FN & 16.42 &  0.00 \\
	Missed detection              & FN & 0.00  & 20.75 \\
	\hline   
	\hline
	Total True                     & T* & 82.95 & 78.31\\
	Total False                    & F* & 17.05 & 21.58\\
	\hline
	\end{tabular}
	}
\end{table}

\begin{figure}[tb]
	\centering
	\disablegraphics{
		\includegraphics[width=.5\linewidth]{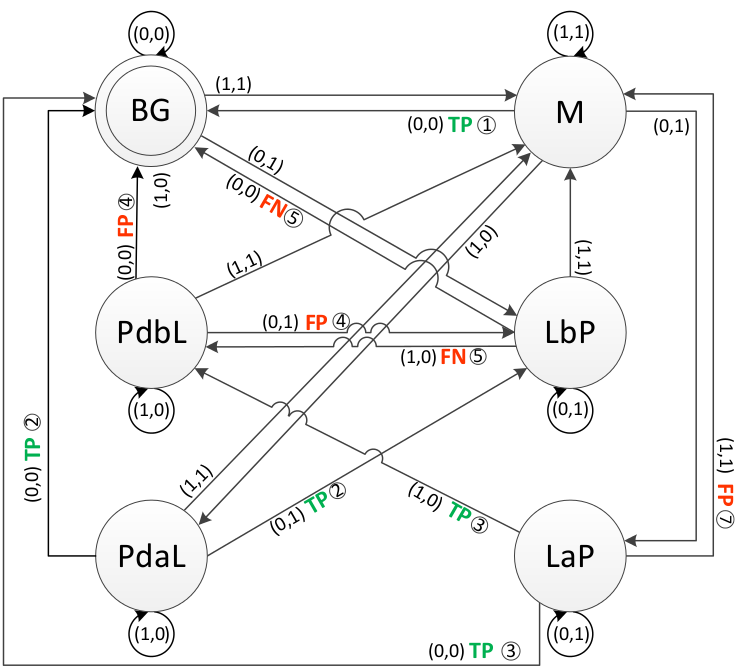}
	}
	\caption{Sequence evaluation state machine.\label{fig:statemachine}}
\end{figure}

As we have seen in the frame-wise classification, most of the pointing misclassifications are non-critial confusions between \enquote{rise}, \enquote{point}, \enquote{fall}, and \enquote{background}.
Accordingly, we can use a window to suppress such misclassifications and obtain a continuous pointing detection.
For this purpose, we use a simple median window to filter the frame-wise detections and eliminate small discontinuities.
To evaluate the resulting continuous blocks of pointing gesture detections, we use a special state machine that is able to distinguish between different types of detection behaviors, see Fig.~\ref{fig:statemachine} and Tab.~\ref{tab:prediction_state_classes}.

As can be seen in Tab.~\ref{tab:prediction_state_classes}, the resulting system exhibits a false positive rate of $0.63\%$.
Furthermore, the system detects all pointing gestures, but unfortunately -- due to the simple filtering mechanism -- it classifies two immediately subsequent pointing gestures into a single pointing gesture detection in $16.42\%$ of the cases.
However, we expect that we can easily improve on these number given a more elaborate frame grouping mechanism than median filtering of the predicted labels.
In $20.32\%$ of the cases, the end frame of the predicted and annotated pointing sequence are exact matches.
The predicted pointing segment is slightly longer than the annotation (i.e., an overlap of \enquote{rise} and/or \enquote{fall} with \enquote{background}) in $23.89\%$ of the detections and slightly shorter in $29.26\%$.
This can be explained with the label ambiguity that we already addressed in Sec.~\ref{sec:evaluation:results:sequence}.

\subsubsection{Baseline}\label{sec:evaluation:results:baseline}

To serve as a baseline, we implemented Nickel and Stiefelhagen's \cite{Nickel2003} \ac{HMM}-based pointing gesture detection system, which -- despite its age -- is still a popular method.
Interestingly, the \acp{HMM} performes bad with our feature set (see Sec.~\ref{sec:features}) and, analogously, the \ac{LDCRF} performs bad with Nickel and Stiefelhagen's features set.
Consequently, we report the \ac{HMM} results based on Nickel and Stiefelhagen's \cite{Nickel2003} original set of features.
As we can see in Tab.~\ref{tab:results:nickel}, the \ac{HMM} correctly detects roughly $95\%$ of frames that belong to rise or fall.
However, this comes at the cost of a substantial amount of false rise and fall detections, especially for \enquote{background} frames but also for \enquote{other} and \enquote{point}.
This leads to the fact that the ability to detect \enquote{point} frames is substantially lower than for the \ac{LDCRF}, see Tab.~\ref{tab:leave-out-subject-trained-with-18-subjects}.
If we consider sequence detection and segmentation based on the \ac{HMM}'s frame-wise detections, see Tab.~\ref{tab:prediction_state_classes}, we notice that the \ac{HMM} does miss entire pointing sequences (\ie, $20.75\%$ missed detection).
Furthermore, we can see that the false positive rate of the \ac{LDCRF} is better compared to the \ac{HMM}'s with $0.63\%$ and $0.83\%$, respectively.


\section{Conclusion and Future Work}\label{sec:conclusion}

We presented how we use \acp{LDCRF} based on depth-based skeleton joint data to perform person-independent pointing gesture detection.
Here, we have shown that \acp{LDCRF} outperform traditional \acp{CRF} and also \acp{HMM} for pointing gesture detection and labeling of individual frames.
Based on the labeled frames of a video sequence, we can use filtering over time to suppress false detections and determine the onset and end of actual pointing gestures.
Thus, segmenting pointing gesture of arbitrary length in video sequences.
This way, we were able to reliably and efficiently detect pointing gestures with a very low false positive rate.
To evaluate our approach, we recorded a novel dataset. 

We leave two important aspects as future work:
First, we intend to improve the pointing sequence extraction based on the frame-wise labeled video frames, specifically to avoid misclassification of two successive pointing gestures as one pointing gesture.
Second, we want to determine the optimal point in time of a pointing gesture to determine the target object, because in preliminary experiments in our dataset we have shown that it has a drastic influence on the ability to correctly determine the pointed-at object (i.e., an improvement of the correct classification rate by up to roughly $10$\% on our dataset).




\bibliographystyle{model1-num-names}
\bibliography{plos}

\end{document}